\documentclass[aps,pre,
   preprint,
    tightenlines,
   superscriptaddress,
   eqsecnum,
   floatfix,
   showpacs,
   preprintnumbers,
   nofootinbib
   ]{revtex4-2}

\usepackage{color}
\usepackage{framed}
\usepackage{subfigure}
\usepackage{time}
\usepackage{graphicx}
\usepackage{enumerate}
\usepackage{latexsym}
\usepackage{bm}
\usepackage{upgreek}
\usepackage{amsthm}
\usepackage{amssymb}
\usepackage{mathrsfs}
\usepackage{amsmath}

\usepackage{physics} 
\usepackage[bottom]{footmisc}
\begin{document}
\preprint{LA-UR-24-24603}
\newcommand{\vphi}{\varphi}
\newcommand{\bq}{\begin{equation}}
\newcommand{\be}{\begin{equation}}
\newcommand{\ba}{\begin{eqnarray}}
\newcommand{\eq}{\end{equation}}
\newcommand{\ee}{\end{equation}}
\newcommand{\ea}{\end{eqnarray}}
\newcommand{\tchi} {{\tilde \chi}}
\newcommand{\tA} {{\tilde A}}
\newcommand{\tq} {{\tilde q}}
\newcommand{\tphi} {{\tilde \phi}}
\newcommand{\tp} {{\tilde p}}
\newcommand{\pstar}{\mbox{$\psi^{\ast}$}}
\newcommand{\sn} { {sn}}
\newcommand{\cn} { {cn}}
\newcommand{\dn} { {dn}}
\newcommand{\rmd}{  {d}}                     
\newcommand{\rme}{  {e}}                     
\newcommand{\Schrodinger}{Schr{\"o}dinger}
\newcommand{\bpsi} {{\bra{\psi}}}
\newcommand{\bphi} {{\bra{\phi}}}
\newcommand{\kpsi} {{\ket{\psi}}}
\newcommand{\kphi} {{\ket{\phi}}}
\newcommand{\calF}{\mathcal{F}}  
\newcommand{\calS}{\mathcal{S}} 
\providecommand{\abs}[1]{\lvert#1\rvert} 
\newcommand{\Braket}[2]{\left\langle{#1}\middle|{#2}\right\rangle}
\newcommand{\br} {\bar{r}}
\newcommand{\bp} {\bar{p}}
\newcommand{\df} {\dot{f}}
\newcommand{\qprod}[2]{ \langle #1 | #2 \rangle} 
\newcommand{\braopket}[3]{\langle #1 | #2 | #3\rangle} 
\interfootnotelinepenalty=10000

\date{\today, \now \ MDT}
\title{Geometric Interpretation of a nonlinear extension of Quantum Mechanics}

\author{Alan Chodos}
\email{alan.chodos@uta.edu}
\affiliation{Dept. of Physics, University of Texas at Arlington, 502 Yates Street, Box 19059, Arlington, TX 76019 }

\author{Fred Cooper} 
\email{cooper@santafe.edu}
\affiliation{The Santa Fe Institute, 1399 Hyde Park Road, Santa Fe, NM 87501, USA}
\affiliation{Theoretical Division and Center for Nonlinear Studies,
   Los Alamos National Laboratory,
   Los Alamos, NM 87545}

\begin{abstract}
We recently introduced a particular nonlinear generalization of quantum mechanics which has the property that it is exactly solvable in terms of the eigenvalues and eigenfunctions of the Hamiltonian of the usual linear quantum mechanics problem.  In this paper we suggest that the two components of the wave function represent the system described by the Hamiltonian $H$ in two different asymptotic regions of spacetime and we show that the non-linear terms can be viewed as giving rise to gravitational effects.
\end{abstract} 
\pacs{3.65Ud, 3.65Ta,
     05.45.-a,  11.10.Lm
          }
\maketitle
\section{\label{s:Intro}Introduction}
In a previous paper \cite{us23} we introduced a new extension of quantum mechanics, in which a pair of state vectors in Hilbert space, $\ket \psi$  and $\ket \phi$, are coupled together non-linearly. The system has the feature that if the underlying linear system is solvable, then the non-linear extension is also solvable.

In this paper, we address a major issue that was left unresolved in \cite{us23}: how to interpret the pair of state vectors that we associate with a dynamical system.  The Schwarzschild solution in general relativity exhibits two distinct asymptotic regions, connected by a non-traversable throat. Inspired by this circumstance, we conjecture that the two state vectors should represent the system in two separate asymptotic regions of spacetime. 
 
To investigate how this might work, in this paper we study the simple example of a particle in $1+1$ dimensions. We choose the Hamiltonian to be that of a free particle, thereby concentrating only on the gravitational effects, which we ascribe to the non-linear couplings in our theory.

Normal general relativity is ``top-down", in the sense that one solves Einstein's equations for the metric and uses that information to determine the geodesics along which particles move. Our approach is ``bottom-up'': we start with the geodesics and use that information to infer the metric. The geodesics, in turn, are determined from suitable expectation values in the underlying extension of quantum mechanics that we have postulated. By requiring that freely falling particles travel on geodesics, we are automatically incorporating a significant element of the equivalence principle.

In a two-dimensional space time, the Einstein tensor vanishes identically. Had we been relying on Einstein's equations, we would have had nothing to work with.  Also, there is no direct analog of the 4-D Schwarzschild solution, since the latter is Ricci flat, and in 2-D if the Ricci tensor vanishes, so does the Riemann tensor, rendering the spacetime flat. 

Nevertheless, by starting with the geodesics, we find non-trivial 2-D metrics, some of which possess the two independent asymptotic regions that we seek. We discover one case that has the exact same singularity and asymptotic structure as the Schwarzschild solution. The metrics we find all have constant curvature. Thus, they are solutions of Jackiw-Teitelboim gravity\cite{Jackiw} \cite{Teitelboim}, a much-studied surrogate for general relativity in two dimensions. We note, however, that the metrics we obtain are not fundamental fields that need to be quantized; they are derived quantities that emerge from the underlying extended quantum mechanical system.

The equations of motion are
\bq \label{eq1.1}
i \frac{\partial}{\partial t} \ket{\psi}  =H \ket{\psi} +g \ket{\phi} \braket{\phi}{\psi}.
\eq
and
\bq \label{eq1.2}
i \frac{\partial}{\partial t} \ket{\phi}  =H \ket{\phi} +g^\ast \ket{\psi} \braket{\psi}{\phi}.
\eq
Here $H$ is the Hamiltonian of the system of interest, whatever that may be, and $g$ is a coupling constant that is in general complex, with $g^* $ its complex 
conjugate. Note that the non-linear terms are universal, in that they are the same regardless of the dynamics described by $H$, unlike in some other recently proposed schemes \cite{Kaplan} .

%

 As shown in \cite{us23} , if we are given an orthonormal pair of solutions of the ordinary \Schrodinger\ equation:
\bq
i \frac{\partial}{\partial t}  \ket{A}  =H   \ket{A}  ,  ~~i \frac{\partial}{\partial t} \ket{B}  =H\ket{B} , ~~\qprod{A}{B} =0, 
\eq
with
\bq
\ket{A} = \sum_n A_n e^{-i E_n t} \ket{ n},~~ \ket {B} = \sum_n B_n e^{-i E_n t} \ket {n},
\eq
then a solution to the non-linear equations is
\ba \label{explicit} 
\ket{\psi} &&= \gamma^{1/2} \left[ e^{i g \omega_0 t}  \sinh \vartheta \ket{A} +  e^{-i g \omega_0 t}   \cosh \vartheta \ket{B} \right] \nonumber \\
\ket{\phi} &&= {\gamma^\ast}^{1/2} \left[ -e^{i g^\ast \omega_0 t}  \sinh \vartheta \ket{A} +  e^{-i g^\star \omega_0 t}   \cosh \vartheta \ket{B} \right] . 
\ea
Here $\gamma= \qprod{\phi}{ \psi}= \omega_0 (\cosh 2 \omega_0 bt)^{i g/b} $, 
 up to an inessential constant phase. The parameter $ b$ is the imaginary part of $g$, which we represent as  $g=a+ib$ and $\omega_0$ and  $\vartheta$ are two additional parameters that characterize the solution, 
 beyond whatever information is resident in the states  $\ket{A}$ and $\ket{B} $.  Note that $ \abs{\gamma}  =\omega_0 ( \cosh 2 \omega_0 b t )^{-1}.$

The derivation of these results, and more details about the properties of the solution, can be found in  \cite{us23}.  What \cite{us23} does not contain, however, is an interpretation of the pair of state vectors that are used to characterize the system of interest.\footnote{The situation is somewhat reminiscent of that facing ordinary wave mechanics in the spring of 1926. At that time, \Schrodinger\  was busy solving his eponymous equation, but he struggled to give an acceptable meaning to the wave function. When Max Born suggested the probability interpretation in the summer of 1926, \Schrodinger\ was repelled, and indeed never reconciled himself to the Copenhagen interpretation, as dramatically expressed in the famous cat experiment that he introduced in 1935.} 

In this theory, there are two state vectors, obeying a pair of coupled non-linear equations, but only one underlying dynamical system, specified by a Hamiltonian $H$. As mentioned above, in this paper we shall explore one possible interpretation that is inspired by some properties of the Schwarzschild solution in general relativity.

As is well-known, the Schwarzschild solution is richer than first appears. When a coordinate transformation is made from the Schwarzschild coordinates to Kruskal-Szekeres coordinates, the apparent singularity at the event horizon disappears, and the spacetime is revealed to have four distinct regions, two of which contain singularities, and two of which are asymptotic, singularity-free regions connected by a non-traversable throat. 

We therefore suggest that   $\ket \psi$  and  $\ket \phi$ can represent the system described by $H$ in two different asymptotic regions of spacetime. A small note of encouragement is that, if one looks at  Schwarzschild coordinates in the two asymptotic regions, time runs in opposite directions. In \cite{us23}, we found that, assuming $H$ is time-reversal-invariant, our non-linear system also possesses a time-reversal invariance that involves interchanging $\ket \psi$   and $\ket \phi$.

In the next section, we pursue some consequences of this suggestion, in the context of a simple example which shows how curvature of space can arise in our non-linear extension of quantum mechanics. 

\section{Geometry from non-linear quantum mechanics} 

Let us consider a system in two spacetime dimensions that consists of a single free particle, by which we mean that the Hamiltonian depends only on the momentum $P$ and not on the position $X$. For any states $\ket A$ and $\ket B$
 evolving according to the usual \Schrodinger\ equation, it then follows that
\bq \label{expx} 
\bra{A(t)} X \ket{B(t)} = \alpha t + \beta.
\eq
(The second time derivative of the left-hand side is the matrix element of the operator $[H,[H,X]] 
$, which vanishes if $H$ depends only on $P$.)

If  $\ket A = \ket B $ , $\alpha$  and $\beta$ must be real parameters. Otherwise, they will in general be complex.

Letting $\ket{ \psi(t)} $ represent our system in some region of spacetime, we can compute the trajectory function

\bq \label{traj1}
X(t) = \braopket {\psi(t)} { X} {\psi(t)} / \qprod {\psi(t)} { \psi(t)} .
\eq 
using equations \eqref{explicit}. Note that  $\braket{\psi(t)}{\psi(t)}= \frac{1}{2} (N+ \tau(t)) $, where  
$ N=2 \omega_0 \cosh( 2 \vartheta)$.  and $\tau(t)  =2 \omega_0 \tanh 2 \omega_0 b t$.
Even for a free particle,  $X(t)$ is no longer simply a linear function. 

From Eq. (\ref{explicit} ), the numerator on the right-hand side of Eq. (\ref{traj1} ) involves the matrix elements $<A|X|A>,, <B|X|B> $and $<A|X|B>$. From 
Eq. (\ref{expx}), the first two each contribute 2 real constants, while the third contributes four. Hence there are a total of eight real constants, so we can write
\bq \label{traj} 
X(t)= \frac{2}{\Delta}  \left[(k_1 t+k_2 )\cosh (b y)+ (k_3 t+k_4 ) \sinh (b y) +(k_5 t+k_6 )  \cos (a y)+(k_7 t+k_8 ) \sin (a y) \right].
\eq
Here  $y=2 \omega_0 t$ and $\Delta =N\cosh (b y)+ 2 \omega_0   \sinh (b y) $.  Note that we have the relation
 \bq
  N\cosh (b y)/\Delta + 2 \omega_0   \sinh (b y)/\Delta =1.
  \eq
so there are really only seven independent constants in this expression.

We suggest that the deviation from linearity can be interpreted as a gravitational effect.
The question we want to answer is, if the trajectory functions defined by Eq. (\ref{traj1}) are geodesics of some metric, what is that metric?
In two spacetime dimensions there are three independent components of the metric, and six of the affine connection. These numbers grow rapidly with dimension so it may be challenging to extend our analysis to higher dimension.
Demanding that the various $X(t)$ determined from Eq. (\ref{traj}) are geodesics for all possible choices of the eight $k_i$ is too restrictive: no metric in 2D will exist. Instead, we ask that they be geodesics for specific choices of the parameters. In this paper we shall examine two possibilities, that we call the one-function and two-function cases. It is not clear to us whether there are any possibilities beyond these two cases for which solutions exist. 

\section{The Problem}
The general problem we face is inverse to the usual one.  In general relativity, typically one is given the metric, either as a solution of Einstein's equation or in some other way.  From the metric one calculates the connection using Christoffel's formula:
\bq \label{Chris} 
\Gamma_{jk}^i=\frac{1}{2}  g^{il }\left(g_{jl,k}+g_{lk,j}-g_{jk,l} \right).
\eq
and then determines the geodesics $x^i(\tau) $ by solving the equation:
\bq \label{geo} 
\frac{d^2 x^i} { d \tau^2} + \Gamma^i_{jk } \frac{dx^j}{d\tau}  \frac {dx^k}{d\tau } = 0.
\eq

Here we are given a set of trajectories  $ \{ x^i(t;k_j) \}  $, where $t$ is the parameter along the particular curve and the $k_j$ label the various curves, and we seek metrics for which these curves are geodesics.
The generic equation for the trajectories in our nonlinear model of quantum mechanics is given by Eq. \eqref{traj}.

Determining a metric from its geodesics is an old problem whose history stretches back to the mists of the nineteenth century.  A modern treatment of this problem has been given by Matveev \cite{Matveev}, which 
also includes a cornucopia of references.  We will follow the procedure outlined in \cite{Matveev}.

In principle, one has to solve the geodesic equation, but  this time for the $\Gamma^i_{jk } $ given the $x^i$.  Then one integrates the compatibility conditions $\nabla_i g_{jk}=0$ to determine the metric. However, there are some complications and subtleties to overcome.

First, we need an extra term in the geodesic equation because the parameter $t$ we are using is not in general the proper time.  The more general form of the equation is:
\bq   
\ddot{x}^i  + \Gamma^i_{jk} \dot{x}^j \dot{x}^k =\hat{g} (\{ \dot{x}^j \})  \dot{x}^i
\eq
where the over dot denotes differentiation with respect to $t$, and $\hat{g}$  is an arbitrary function that encodes the relation between $t$ and the proper time $\tau$. We shall assume that it is possible to choose coordinates such
that $t=x^0$.  The  $i=0$ component of the geodesic equation reads
\bq
\hat{g} = \Gamma^0_{00} + 2 \sum_{j \neq 0} \Gamma^0_{j0} \dot{x}^j + \sum_{j,k \neq 0} \Gamma^0_{jk} \dot{x}^j \dot{x}^k.
\eq
We insert this expression into the remaining equations to obtain, for $i \neq 0$,
\bq \label{geoeq}
\ddot{x}^i + \Gamma^i_{00} + 2 \sum_{j \neq 0} \Gamma^i_{j0} \dot{x}^j + \sum_{j,k \neq 0}  \Gamma^i_{jk} \dot{x}^j \dot{x} ^k = \dot{x}^i \left[\Gamma^0_{00} + 2 \sum_{j \neq 0} \Gamma^0_{j0} \dot{x}^j 
+ \sum_{j,k \neq 0} \Gamma^0_{jk} \dot{x}^j \dot{x}^k \right].
 \eq
 In 2D this equation becomes:
 \bq \label{geod2D} 
 \ddot{x}^1 + (2 \Gamma^1_{10} - \Gamma^0_{00})  \dot{x}^1 +(\Gamma^1_{11}-2 \Gamma^0_{10} )(\dot{x}^1)^2 - \Gamma^0_{11} (\dot{x}^1)^3+\Gamma^1_{00} =0.
 \eq

%

It would be nice if, given a sufficient number of trajectories $ x^i$, one could completely determine the connection coefficients  $\Gamma_{jk}^i.$ But that is never the case, because the geodesic equations possess a ``gauge invariance",  $\Gamma_{jk}^i 
\rightarrow \Gamma_{jk}^i- \delta_j^i \phi_k-\delta_k^i \phi_j$, where the $\phi_j $are arbitrary. This transformation will only change the value of the function $\hat{g}$, leaving the form of the geodesic equation unaltered.

In 2D there are 6 components of the connection, and the geodesic equations will determine the four gauge-invariant combinations $\Gamma_{11}^0,  ~(2 \Gamma^1_{10} - \Gamma^0_{00}),~\Gamma^1_{00} , ~(\Gamma^1_{11}-2 \Gamma^0_{10} )$ , leaving the remaining two arbitrary. The compatibility equations $\nabla_i g_{jk}=0 $ are not gauge invariant, so they require more input than we seem to have available. 

One can circumvent this difficulty by working with auxiliary functions, related to the metric, that are gauge invariant. In 2D, these can be taken to be
\bq
a_{ij} =|Det (g )|^{-2/3} g_{ij}. 
\eq

The $a_{ij} $ obey the following equations:
\ba \label{aeq}
\partial_t a_{00}+ 2 K_0 a_{01}  - \frac{2}{3}  K_1~ a_{00}=0 , \nonumber \\
2 \partial_t a_{01} + \partial_x a_{00} + 2 K_0 a_{11} +\frac{2}{3}  K_1  a_{01}- \frac{4}{3}K_2 a_{00} =0, \nonumber \\
\partial_t  a_{11}+2 \partial_x a_{01}+\frac{4}{3}  K_1 a_{11}- \frac{2}{3}  K_2 a_{01} - 2 K_3 a_{00} =0, \nonumber \\
\partial_x a_{11} +\frac{2}{3}K_2 a_{11}- 2 K_3 a_{01}  =0,
\ea
where 
\bq
K_0=- \Gamma^1_{00},~~ K_1=\Gamma^0_{00}- 2 \Gamma^1_{01},~~K_2= - \Gamma^1_{11}+2 \Gamma^0_{01},~~
K_3= \Gamma^0_{11}.
\eq
The procedure is to solve these equations for the  $a_{ij}$, and then to determine the $g_{ij} $ from
\bq \label{met} 
g_{ij}=\frac{a_{ij}} {|Det(a)|^2 }.  
\eq.

One can then solve for all the  $\Gamma_{jk}^i $ and verify that the input information conveyed by 
the  $ K_i$ is reproduced. Analogous equations exist in higher dimension, but in the analysis to follow we shall concentrate on 2D for simplicity.

\section{Solution of the one function Case}
In this section we consider what we call the one-function case, in which we choose a particular combination $f(t)$ of the general 
$X(t)$ with fixed $k_i$ (for example, all $k_i$ vanish except one), and then demand that any multiple of that combination be a geodesic.  In section VI,  we shall consider the two function case which will involve two combinations of $f_1(t)$ and $f_2(t)$.

From Eq. (\ref{geod2D} ) we obtain the geodesic equation for $x^1 =f(t)$
\bq \label{geodesic}
\ddot{f} +\Gamma^1_{00} + 2 \Gamma^1_{10} \dot{f} + \Gamma^1_{11} \dot f^2 =\dot{f} (\Gamma^0_{00}+ 2 \Gamma^0_{10} \dot{f} + \Gamma^0_{11} \dot f^2)
\eq
Since $f$ contains an arbitrary multiplicative constant, we require that the  coefficients of powers of  $ f$  must vanish separately. This leads to 
\bq
\Gamma^1_{00}=0; \Gamma^1_{11}- 2 \Gamma^0_{10}=0,~~ \Gamma^0_{11} =0; ~\Gamma^0_{00}- 2 \Gamma^1_{10}=\frac{\ddot{f}}{\dot{f}}~,
\eq
so that
\bq \label{geodesic2}
\ddot{f} + ( 2 \Gamma^1_{10} -\Gamma^0_{00}) \dot{f}  = 0  .\eq
It will be useful to define 
\bq
h(t) =\frac{\ddot{f}}{\dot{f} }   
\eq
For our problem, the  $K_i$ used in Matveev (Eq. 17 of Ref. \cite{Matveev})  are given by:
\bq
K_0=0; ~~ K_1=h(t); K_2=K_3=0.
\eq
From Eq. (\ref{aeq}   ), we obtain the Liouville \footnote{Note that this is R. Liouville, not his more famous namesake J. Liouville, after whom the Liouville equation is named.} system  \cite{Liouville} :

\ba \label{Liouville}
\partial_t a_{00}  - \frac{2}{3}  h(t)~ a_{00}=0 , \nonumber \\
2 \partial_t a_{01} + \partial_x a_{00} +\frac{2}{3}  h(t) a_{01}=0, \nonumber \\
\partial_t a_{11}+2 \partial_x a_{01}+\frac{4}{3}  h(t) a_{11}=0, ~~
\partial_x a_{11}=0.
\ea
It  follows that $a_{11}$  is independent of $x$, $a_{01}$ is at most linear in  $x$, and $a_{00}$ is at most quadratic in $x$. So we write
\bq \label{aexp}
a_{11}=p_0 (t); ~a_{01}=q_0 (t)+q_1 (t) x;~ a_{00} =r_0 (t)+r_1 (t)x+r_2 (t)x^2.
\eq

We see that only the gauge invariant combinations enter into Eqs. (\ref{Liouville}). Thus one only needs $h(t)$ to solve for the $g_{ij}$ and then one can determine
all six nonzero  $\Gamma^i_{jk}$ directly from the $g_{ij}$.

Once we determine the $a_{ij}$ the $g_{ij}$ are given by:

\bq \label{gij} 
g_{ij} = \frac{a_{ij}}{ |Det(a)| ^2}
\eq

Inserting Eq.(\ref{aexp})  into the equations for $a_{ij}$, and equating the coefficients of each power of $x$ separately to zero, we obtain six equations, which divide into a single equation for $r_0(t)$, a pair of equations for $r_1 (t)$ and  $q_0 (t)$, and three equations for $r_2 (t), q_1 (t)$, and $p_0 (t)$. They are:
\ba
&&\dot{r} _0- \frac{2}{3}  h(t) r_0=0 ,\nonumber \\
&&\dot{r} _1-\frac{2}{3} h(t) r_1=0 ; ~~ 2 \dot{q}_0+r_1+\frac{2}{3} h(t) q_0=0, \nonumber \\
&&\dot{r} _2-\frac{2}{3} h(t) r_2=0 ;~\dot{p}_0+\frac{4}{3} h(t) p_0+2q_1=0 ;~\dot{q} _1+\frac{2}{3}  h(t) q_1+r_2=0.
\ea
Now $h(t)=  \frac{\ddot{f}} { \dot{f}} $ which is of the form $\frac{\dot{g}}{g}$.  Since for any function $s(t)$ and any constant $\alpha$, 
\bq  \label{iden}
\dot{s} +\alpha \frac{\dot{g}}{g}  s= g^{-\alpha}  \frac{d}{dt}  (g^{\alpha} s),
\eq
 we immediately obtain
\bq
r_i (t)= \bar{r}_i (\dot{f})^{2/3},     i=0,1,2
\eq
where the $\bar{r}_i $ are constants. We can proceed to integrate the remaining equations in terms of three more constants of integration $c_i$. We then find for the $a_{ij}$
\ba
a_{00} &&= \dot{f} (t)^{2/3} \left(\bar{r}_0+\bar{r}_1 x+\bar{r}_2 x^2\right) \nonumber \\
a_{01} &&= \left(  -\frac{c_1 \bar{r}_1}{2}- c_2 \bar{r}_2 x- f(t)
   (\frac{\bar{r}_1}{2}+\bar{r}_2 x ) \right) /  \dot{f} (t)^{1/3}. \nonumber \\
   a_{11} &&= \bar{r}_2  (c_3+2 c_2 f(t)+f^2 (t) )/ \dot{f} (t) ^{4/3}.
\ea

One can simplify the expressions for $a_{ij}$ by the change of variables:
\bq
x \rightarrow X- \alpha;    f(t) \rightarrow T-\beta
\eq
We can eliminate the linear dependence on $x$ in $a_{00}$ and the linear dependence of $a_{11}$ on $f(t)$  by choosing 
\bq
\alpha= \frac{\br_1} {2 \br_2} ; ~~ \beta= c_2.
\eq
We then have that
\ba
a_{00}&&= \br_2 (a'+X^2) \dot{f} (t)^{2/3},~~
a_{01}= -\frac{\br_2 (b' +T X)}{\dot{f} (t)^{1/3} } \nonumber \\
a_{11} &&= \br_2  \frac{(T^2 +c')} {\dot{f} (t)^{4/3} }
\ea
where 
\bq
a'=\frac{ \br_0}{\br_2 }- \frac{\br_1^2}{4 \br_2^2} ,
b'= \frac{(c_1-c_2) \br_1}{2 \br_2},~~
c'= c_3-c_2^2.
\eq

Calculating the determinant we find
\ba
Det[a] && =  \frac{ \br_2^2 (-b'^2 + a'(c'+T^2)- 2 b' T X+ c' X^2)}{ {\dot{f} (t)}^{2/3} } \nonumber \\
&&=  \frac{ \br_2^2 [-(b' +T X)^2+ (c'+T^2) (a'+X^2)]} { {\dot{f} (t)}^{2/3}  }
\ea
The metric in the original coordinates $(t, x=f(t))$ can be written as 
\bq
g_{\alpha \beta} = \frac{1}{\br_2^3 D^2} \left(
\begin{array}{cc} 
(a'+X^2 )\dot{f} (t)^2 &~~- (b'+T X )\dot{f} (t) \\
-(b'+TX) \dot{f} (t)  &~~  c'+T^2 \\
\end{array}
\right)
\eq 
where
\bq
D=-  (b'+T X)^2+ (c'+T^2) (a'+X^2).
\eq
Now we have in the original coordinates $x^\alpha$
\bq
ds^2 = g_{\alpha \beta} dx^\alpha d x^\beta
\eq
now since
$ T=f(t) +c_2$, 
\bq
dT= \dot{f} (t) dt ;   dX=dx
\eq
If we change coordinates to  ${ T, X}$ , we have due to the invariance of $ds^2$ the factors of $\dot{f} (t)$ in $g_{\alpha \beta}$  
get absorbed into the definition of $dT$.
Letting  $X^\alpha = \{T,X \} $ one has
\bq
ds^2= h_{\alpha \beta} dX^\alpha dX^\beta
\eq
where now 
\bq
h_{\alpha \beta} = \frac{1}{\br_2^3 D^2} \left(
\begin{array}{cc} 
(a'+X^2 ) &~~ -(b'+T X ) \\
-(b'+TX)   &~~  c'+T^2 \\
\end{array}
\right).
\eq 

The inverse metric is given by
\bq
h^{\alpha \beta}=\br_2^3 D \left(
\begin{array}{cc}
(c'+T^2)  &~~  (b'+TX ) \nonumber \\
(b'+T X ) &~~  ( a'+X^2) \ \\
\end{array}
\right).
\eq
The nonzero connections calculated from Eq. (\ref{Chris} )  are given by:

\begin{eqnarray}
 \Gamma^0_{00}  &&= 2 \frac{(b'X -a'T)}{D};~~    
\Gamma^0_{10}  = \frac{(b'T -c'X)}{D}    \nonumber \\
\Gamma^1_{10} &&= \frac{(b'X -a'T)}{D} ;~~
\Gamma^1_{11}  = 2 \frac{(b'T -c'X)}{D}. \\
\end{eqnarray}

Note that in this coordinate system both $ K_1$ and $K_2$ are zero.  Previously $K_1= - \frac{\ddot{f}}{\dot{f}}$. \\
The Ricci Curvature is given by
\bq
\frac{1}{D^2} 
\left(
\begin{array}{cc}
 \left( {a'}+X^2\right) \left( {a'}
    {c'}- {b'}^2\right) &~~
  \left( {b'}^2- {a'}  {c'}\right) (b'+ T X) \\
\left( {b'}^2- {a'}  {c'}\right)(b'+T X)   &~~
   - \left( {c'}+T^2\right) \left( {b'}^2- {a'}
    {c'}\right) \\
\end{array}
\right)~.
\eq
We can factor out $(a'c'-b'^2)$ to obtain 
\ba
R_{\alpha \beta}&& = \frac{a'c'-b'^2 }{ D^2} \left(
\begin{array}{cc} 
(a'+X^2 ) &~~ -(b'+T X ) \\
-(b'+TX)   &~~  c'+T^2 \\
\end{array}
\right) \nonumber \\
&&= (a'c' -b'^2) \br_2^3 h_{\alpha \beta}.
\ea
We see that the Ricci curvature is proportional to the metric $h$ as it must be in two dimensions.
The scalar curvature is given by
\bq
R= 2 \br_2^3 \left(a'c'-b'^2\right).
\eq

\section{	Geometry of the One-function Solution}

In the previous section, we determined the metric in the one-function case, in terms of six constants of integration. One of these is an inessential overall constant, which we choose arbitrarily. Two of them are absorbed in a change of variables from the original $x$ and $t$ to $X$ and $T$ :
\bq
X=x+\alpha  ;   T=f(t)+ \beta
\eq

Here, by construction, the geodesics are $x=kf(t)+k_0 $, or, what is the same thing, 
\bq
X= kT+X_0,
\eq
i.e. straight lines in the $(X,T)$ plane. The remaining 3 constants play a significant role in determining the metric, which, in$ (X,T) $ coordinates, has the form:
\bq
h_{\alpha \beta}= \frac{1}{ \br_2^3D^2 } \left( \begin{array}{cc} 
 p_{00} & p_{01} \\
 p_{10} &p_{11} \\
 \end{array}
 \right) ,
\eq
with  $p_{00}= X^2+a'$ ;    $p_{01}=p_{10}=-(XT+b')$   ; $ p_{11}= T^2+c'$. 
Here $D= p_{00} p_{11}-p_{01}^2=a'T^2+c' X^2-2b'XT+a'c'-b'^2$.

Setting $\bar{r}_2=1$, the Ricci scalar $R=2(a' c'-b'^2)$, which is twice the Gaussian curvature.  Our space is one of constant curvature, which can be negative, positive or zero depending on the choice of parameters.

We see that $Det |h|= D^{-3} $, so the curve $D=0$, along which  $h_{\alpha \beta }$ is singular, separates a region of Euclidean signature from one of Minkowski signature. No singularity exists if we choose $a' > 0$, $ c' > 0 $ and $a' c'-b'^2 \ge 0$, in which case the entire plane has Euclidean signature. Since we are interested in spacetime, not Euclidean space, we consider cases for which $D$  can vanish. We shall regard the curve $D=0 $ as separating the physical space $D<0$ from the $D>0$ region, which we take to be unphysical (although not everyone agrees; see \cite{Bars}).

\begin{figure}
\includegraphics[width=0.4 \linewidth]{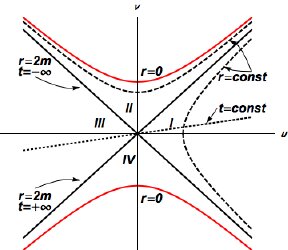} 
\caption{Regions of spacetime associated with the Schwarzschild geometry using the Kruskal coordinates $u,v$. The angular coordinates have been suppressed. The dotted lines shown in regions I and II are typical curves of constant $r$. Trajectories of constant $t$ are straight lines through the origin. The solid black lines are the event horizon $r=2m.$ The red hyperbola in regions II and IV is the singularity at $r=0$. The 2D spacetime discussed in Section V maps exactly onto this picture, although in the Schwarzschild case the geodesics are not straight lines.} 
\label{geometry}
\end{figure}

To proceed, we choose our parameters to bring our 2D model into a form similar to that used in a standard analysis of the radial geodesics of the Schwarzschild metric  \cite{Fuller and Wheeler}. We take $\br_2=1$,  $a'=c'=0$ and $b'<0$, and furthermore we make a linear change of variables:
$T=v+u  ;X=v-u$ in terms of which the spacetime interval becomes
\bq
 ds^2=\frac{2}{D^2} \left[(2u^2+|b'|) dv^2-(4u v) ~du dv+(2v^2-|b'|) du^2 \right]
\eq
with $D=|b'|[2(v^2-u^2 )-|b'|]$ .Thus the singular curve is a hyperbola, as shown in Fig. (\ref{geometry})
We write a typical geodesic as $v=\xi (u-u_0 )$.  Along the geodesic, $dv= \xi du$,  which implies

\bq
ds^2=\frac{2K}{D^2} ~ du^2,
\eq
where $ K=\xi^2 (2u_0^2+|b'|)-|b'| $.
Thus, we find that the condition for a null geodesic is
\bq
\xi^2= \frac{\abs{b'} }{2u_0^2+\abs{b'}}
\eq

The condition that a geodesic be tangent to the hyperbola $(v^2-u^2 )=(|b'|)/2$ is 
\bq
\xi=\frac{dv}{du}= \frac{ u}{v},
\eq
which, after a bit of algebra, is seen to be the same as the condition for a null geodesic. That is, all null geodesics are tangent to the singular hyperbola, and all lines tangent to the singular hyperbola are null.  The parameter $\xi$, which measures velocity (actually inverse velocity, since we are regarding $ v$ as the time-like parameter) depends on the particular null geodesic, reaching its maximum of 1 for $u_0=0$.

Through any point, one can draw 2 null geodesics. In regions I and III of the figure, one of these is tangent to the upper branch, and one to the lower. In region II, both are tangent to the upper branch, and in region IV, both are tangent to the lower branch. The origin $(u,v)=(0,0)$ is special, in that each of the two null geodesics is asymptotically tangent to both the upper and lower branches. 

The time-like geodesics are those with $K>0$.  Inside a typical light cone in regions I and III, there will be some geodesics that intersect both the past and future singularities; there will be others that originate in the past singularity but escape to infinity in the future, and there will be still others that come from infinity in the past and intersect the future singularity. There are no time-like geodesics that escape to infinity in both the past and the future, which is the same as saying there are none that travel between regions I and III. The throat is not traversable for time-like geodesics. 

On the other hand, space-like ($K<0$) geodesics that intersect the line segment $(0,v)$ with $-|b'|/2<v<(|b'|)/2 $ do travel through the throat between regions I and III.

If we take the view that the  ``white hole''  singularity is not physical because a black hole should form from non-singular data in the past, we can concentrate on the part of the spacetime with $v>0$.  Then the special null geodesics through the origin act as an event horizon, since all time-like geodesics in region II end in the singularity, whereas in regions I and III there are at least some time-like geodesics that can escape to infinity.

\section{Two different two-function cases} 

The one-function case has the advantage of simplicity, but the disadvantage that the function itself can be eliminated from the metric by a change of coordinates, so that all one-function metrics are essentially the same. They are governed by the parameters we have called $a',b',c'$,  but these are constants of integration that have nothing to do with the function we started with.
It is therefore of interest to examine more complicated cases. Given the constraints imposed by the geodesic equation, it is unlikely that one can make use of the full set of independent functions inherent in the trajectory of Eq. (\ref{traj}). However, there are at least two ways to introduce a pair of functions, that we now describe.

\subsection{case A}  
The simplest way to introduce a second function is to choose the set of trajectories to be of the form
\bq
X(t)=kf_1 (t)+f_2 (t)
\eq
and require these to be geodesics for all values of the parameter $ k.$ We insert this form into Eq.(\ref{geoeq}), and equate powers of $k$. We find the conditions
\bq \label{gauge1}
\Gamma_{11}^0=0; ~\Gamma_{11}^1- 2\Gamma_{01}^0=0; ~\Gamma_{00}^0-2\Gamma_{01}^1=\frac{\ddot{f_1} }{ \dot{f} _1 }; ~\Gamma_{00} ^1=\frac{w}{\dot{f} _1} ,
\eq
So that $K_3=K_2=0$. $K_1= \frac{\ddot{f_1} }{ \dot{f} _1} $ and $K_0 = - \frac{ w }{ \dot{f} _1 }$. 
where we have defined  $w=\ddot {f}_1 \dot{f}_2 -  \ddot {f}_2 \dot{f}_1 $. We can then use these as input to Eq. (\ref{aeq}) for the quantities  $a_{ij}$ . 
The $a_{ij} $ equations reduce to: 
\ba \label{aeq3}
&&\partial_t a_{00}- 2 K_0 a_{01}  - \frac{2}{3}  K_1~ a_{00}=0 , \nonumber \\
&&2 \partial_t a_{01} + \partial_x a_{00} + 2 K_0 a_{11}  =0, \nonumber \\
&&\partial_t  a_{11}+2 \partial_x a_{01}+\frac{4}{3}  K_1 a_{11} =0, \nonumber \\
&&\partial_x a_{11}   =0.
\ea

As in the one-function case, $a_{11}$  is independent of $x$, $a_{01}$ is at most linear in $x,$ and $a_{00}$  is at most quadratic in $x$. We can therefore use the same parameterization as in the one-function case Eq.(\ref{aexp}), but of course the equations for the six functions of time will be different.  We find the set of equations:
\ba \label{eq2funa}
 \dot{r}_2 - \frac{2}{3} \frac{\ddot{f} _1}{\dot{f}_1}  r_2=0; ~~\dot{q}_1 +\frac{1}{3}  \frac{\ddot{f} _1}{\dot{f}_1}  q_1+r_2=0, \nonumber \\
 \dot{p}_0 + \frac{4}{3} \frac{\ddot{f} _1}{\dot{f}_1}  p_0+2 q_1= 0; ~~\dot{r}_1-\frac{2}{3}  \frac{\ddot{f} _1}{\dot{f}_1}  r_1-
 \frac{2}{\dot{f}_1} w q_1=0, \nonumber \\
 2 \left(\dot{q}_0 + \frac{1}{3} \frac{\ddot{f} _1}{\dot{f}_1}  q_0 \right)+ r_1- \frac{2}{\dot{f}_1} w p_0= 0; ~~\dot{r}_0-\frac{2}{3}  \frac{\ddot{f} _1}{\dot{f}_1}  r_0-
 \frac{2}{\dot{f}_1} w q_0 =0, \nonumber \\
\ea 
If we set $f_2=0$, we recover the equations for the one-function case, as expected.

We define the functions $u,v, w$ by 
\bq
u(t) = \frac{\dot{f}_2} {\dot{f}_1} ,~~ v= \dot{f}_1 f_2- \dot{f}_2 f_1,  ~w=\ddot {f}_1 \dot{f}_2 -  \ddot {f}_2 \dot{f}_1.
\eq
We see that we can relate $f_2, f_1,u,v$ using
$f_2= u f_1 + \frac{v} {\dot{f}_1}$, or $u f_1-f_2 = - \frac{v}{\dot{f}_1}$.
Using Eq. (\ref{iden}), we can solve these equations sequentially starting with the equation for $r_2$.  
We find the following functional forms for the 6 independent functions satisfy Eq. (\ref{eq2funa}):
\ba \label{soleq2fun}
 r_2(t) &&= \bar{r}_2 ( \dot{f}_1)^{2/3} , \nonumber \\
 q_1(t)&&= ( \dot{f}_1)^{-1/3} (\bar{q}_1 - \bar{r}_2 f_1(t)) , \nonumber \\
 p_0(t)&&= ( \dot{f}_1)^{-4/3}(\bar{p}_0 - 2 \bar{q}_1 f_1(t) + \bar{r}_2 f_1(t)^2), \nonumber \\
 r_1(t)&&= ( \dot{f}_1)^{2/3}(\bar{r}_1 - 2 \bar{q}_1 u(t) -2 \bar{r}_2 \frac{v(t)}{ \dot{f}_1} ) , \nonumber \\
q_0(t)&&=  \frac{1}{2} ( \dot{f}_1)^{-1/3} \left[ \bar{q}_0+ 2 \bar{r}_2 v(t) \frac{f_1}{\dot{f}_1} - 2 \bar{p}_0 u(t) -\bar{r}_1 f_1 +2\bar{q}_1( u(t) f_1(t) - v(t) /\dot{f}_1) \right] ,\nonumber \\
  r_0(t)&&=  ( \dot{f}_1)^{2/3} \left[ \bar{r}_0- \bar{q}_0 u(t) + \bar{r}_2 ( \frac{v(t)} {\dot{f}_1})^2  +\bar{p}_0 u^2(t) -\bar{r}_1 v(t)/\dot{f}_1+2 u(t) \bar{q}_1 v(t)/\dot{f}_1 \right] . \nonumber \\
\ea
In terms of this solution we have Eq. (\ref{aexp})
\bq 
a_{11}=p_0 (t); ~a_{01}=q_0 (t)+q_1 (t) x;~ a_{00} =r_0 (t)+r_1 (t)x+r_2 (t)x^2.
\eq
We find that $ |Det(a)|$ can be written as:
\ba
&&|Det(a)|=\frac{ 1} {4( \dot{f}_1)^{2/3}}  \times \nonumber \\
&&\left[e_2 - 4 e_4 x - 4 e_1 x^2 - e_3 f_1(t) ^2 + 4 (e_4 + 2 e_1 x) f_2(t) - 
 4 e_1 f_2(t)^2 + 2 f_1(t)  (e_5 - 2 e_6 x + 2 e_6 f_2(t))\right]. \nonumber \\
 \ea
where
\ba
&&e_1=\bar{q}_1^2-\bp_0 \br_2,~ e_2=-\bar{q}_0^2 + 4 \bp_0 \br_0, ~e_3 = \br_1^2 - 4 \br_0 \br_2,~ e_4=\bar{q}_0 \bar{q}_1 - \bp_0 \br_1 \nonumber \\
&&e_5=-4\bar{q}_1 \br_0 + \bar{q}_0 \br_1, ~e_6= \bar{q}_1 \br_1 -\bar{q}_0 \br_2.
\ea
The metric is given by Eq. (\ref{gij}). 

The affine connections satisfy the gauge invariant conditions Eq. (\ref{gauge1}).  
The Ricci curvature has the property that 
\bq
R_{\mu \nu} = \frac{1}{2} R g_{\mu \nu}.
\eq
where the scalar curvature $R$ is a constant and is given by
\ba \label{ricci2}
R&&= 2 \bar{r}_0 
\left(\bar{p}_0 \bar{r}_2 -(\bar{q}_1)^2 \right)  + \bar{q}_0 \bar{q}_1 \bar{r}_1 - \frac{1}{2} \left(\bar{p}_0 (\bar{r}_1)^2 + \bar{r}_2 (\bar{q}_0)^2 \right) \nonumber \\
&&= - 2 \br_0 e_1 + \frac{1}{2} \br_1 e_4 + \frac{1}{2} \bar{q}_0 e_6.
\ea
We can re-express the metric by changing the time coordinate from $t$ to $T$ via
\bq
f_1 (t)=T-T_0.
\eq
Just as with the one-function case, this will absorb the pre-factors of $\df_1(t)$. In addition, we have the relation 
\bq
\frac{\df_2}{\df_1} = \frac{df_2(T) }{dT} =f'_2 (T),
\eq
 which also implies that 
 \bq
\frac{v} {\df_1} =f_2 (T)-(T-T_0) f'_2 (T)
\eq
Using these, we find that the metric now depends on $ f_2 (T)$ and $ f'_2 (T)$, as well as explicitly on the coordinates $ x$ and $T$. As we see from Eq. (\ref{ricci2}), the Ricci scalar is a constant, so it does not provide evidence as to whether the dependence on $f_2$ has real geometrical significance or is merely a coordinate artifact.

\subsection {case B } 
A second way to involve two functions is to demand that
\bq \label{x1} 
x^1(t)=\kappa_1 f_1 (t)+\kappa_2 f_2 (t)
\eq
is a geodesic for arbitrary choice of the constants $\kappa_i$. At first, this does not seem possible, because the terms linear in the $\kappa_i$  in the geodesic equation impose the condition
\bq
(\kappa_1\dot{f}_1 + \kappa_2 \dot{f}_2) (\Gamma^0_{00} - 2 \Gamma^1_{10}) = \kappa_1 \ddot{f}_1 + \kappa_2 \ddot{f}_2.
\eq
which can hold for all $\kappa_i$ only if $\ddot{f}_1 / \dot{f}_1 =\ddot{f}_2 / \dot{f}_2$ . But that implies
 $\dot{f}_2= \beta \dot{f}_1$  for some constant $\beta$ , and hence $f_2 (t)=\beta f_1 (t)+\beta_0$, so we are essentially back to the one-function case.
However, this ignores the possibility that the  $\Gamma_{jk}^i $can depend on $x $ as well as on $t.$ We rewrite
Eq.(\ref{x1}) as
\bq
\kappa_2=\frac{x-\kappa_1 f_1 (t)} { f_2 (t)},
\eq
where we now treat $x$ as the coordinate, not the given function of $t$ (this should be valid as long as we are on the geodesic). We substitute this expression into the geodesic equation and equate powers of  $\kappa_1$.

The terms cubic and quadratic in $\kappa_1$ lead as before to the conditions
\bq
\Gamma_{11}^0=0 ; \Gamma_{11}^1-2 \Gamma_{01}^0=0.
\eq
But the terms linear in $\kappa_1$ and independent of $\kappa_1$ yield new conditions, which can be solved and lead to:
\bq
\Gamma_{00}^0-2 \Gamma_{10}^1= \frac{\ddot{f}_1 f_2- \ddot{f}_2 f_1} {\dot{f}_1 f_2- \dot{f}_2 f_1 }
\eq
and
\bq
\Gamma_{00}^1= (\frac{\ddot{f}_1 \dot{ f}_2- \ddot{f}_2  \dot{f}_1} {\dot{f}_1 f_2- \dot{f}_2 f_1 })x   
\eq
Letting  
\bq
v= \dot{f}_1 f_2- \dot{f}_2 f_1, ~~  w=\ddot{f}_1 \dot{f}_2- \ddot{f}_2 \dot{f} _1,
\eq  
we have
\bq \label{gauge} 
\Gamma_{00}^0-2 \Gamma_{10}^1= \frac{\dot{v}} {v}  ;~~\Gamma_{00}^1 = \frac{w}{v} x.
\eq

Using this information, we find that, once again, $a_{11} $ is independent of $x$, $a_{01}$ is at most linear in $x$, and $a_{00}$ is at most quadratic in $x$. So we can continue to use the same parameterization of the $a_{ij}$ in terms of  $p,q,r$,   Eq.(\ref{aexp}). In this case, we find the equations
for the $a_{ij}$ , i.e. Eq.( \ref{aeq}) can be written as:
\ba \label{pqreq} 
&&\dot{r}_0- \frac{2}{3} \frac{\dot{v} }{v} r_0=0; \nonumber \\
&&\dot{r} _1- \frac{2}{3}  \frac{\dot{v} }{v} r_1-2  \frac{w}{v}  q_0=0 ;~~  \dot{q}_0+\frac{1}{3}  \frac{\dot{v} }{v} q_0+\frac{1}{2} r_1=0; \nonumber \\
&&\dot{r}_2-\frac{2}{3}  \frac{\dot{v}}{v} r_2- 2 \frac{w}{v} q_1=0;~ \dot{q}_1+ \frac{1}{3}  \frac{\dot{v} }{v}q_1+r_2-\frac{w}{v} p_0=0; \nonumber \\
&& \dot{p}_0+\frac{4}{3}  \frac{\dot{v} }{v}p_0+2q_1=0.
\ea
 If we set $w=0$,  then $f_2 (t)=\beta f_1 (t)+\beta_0$, and hence $v=\beta_0 \dot{f}_1$, so these equations reduce to the one-function case.
 
We can solve the equations for $r_1$ and $q_0$  by first  introducing new variables which scale out the $v$ dependence:
\bq
 A_1= v^{1/3}  q_0 ; B_1 = v^{-2/3} r_1
 \eq.
 Then we have  that $A_1$ and $B_1$ satisfy the equations  
 \bq
 \dot{A}_1+ \frac{1}{2} v B_1=0;~~\dot{B}_1 - \frac{2 w}{v^2} A_1=0.
 \eq
 Making the ansatz
  \bq
A_1=  \calF(t) =  \left(\alpha f_1 +\beta f_2 \right),
  \eq
  where $\alpha$ and $\beta$ are arbitrary constants, one finds immediately that
  \bq
  B_1=-2  \frac{\dot{\calF}}{v}
  \eq
The ansatz for $\calF$ satisfies the equation:
\bq
\ddot{\calF} v - \dot{\calF} \dot{v} + \calF w = 0.
\eq
Thus we have
\bq
q_0 = v^{-1/3}  \left(\alpha f_1 +\beta f_2 \right), r_1= -2 v^{-1/3}  \left(\alpha\dot{f}_1 +\beta \dot{f} _2 \right)
\eq
%
To solve for the remaining three variables we again rescale $p_0$ , $r_2$ and $q_1$ by introducing:
\bq
A_2= v^{4/3} p_0; B_2 = v^{-2/3} r_2; C= v^{1/3} q_1.
\eq

We now have the rescaled equations
\bq
\dot{A_2}+ 2 v C =0; ~~ \dot{B}_2- \frac{2 w}{v^2} C=0; ~~ \dot{C}+ v B_2 - \frac{w}{v^2} A_2=0.
\eq

The variables $A_2,B_2,C$ satisfy the constraint.
\bq \label{constraint} 
\frac{d}{dt} \left(A_2B_2-C^2 \right)=0.
\eq
Making the assumption that $A_2, B_2$  and $C$  depend on $t$ only through  the functions $f_1,f_2, \dot{f}_1,\dot{f_2}$ one finds the solution
\ba
A_2 &&= \left( m_1 f_1^2 + m_2 f_1 f_2 + m_3 f_2^2 \right)  \nonumber \\ 
B_2 &&= \frac{1}{v^2}  \left(m_1  \dot{f_1} ^2 +m_2   \dot{f}_1 \dot{f}_2 + m_3  \dot{f_2} ^2 \right) \nonumber \\
C&&=- \frac{1}{v}  \left( m_1  f_1 \dot{f}_1 +m_3 f_2 \dot{f}_2 +\frac{1}{2} m_2 (\dot{f}_1 f_2+\dot{f_2} f_1)  \right).
\ea

Here $m_1,m_2$ and $m_3$ are the three arbitrary constants of the solution.  They enter into the solution of  the constraint equation  Eq. (\ref{constraint})
\bq
A_2 B_2-C^2 =  m_1 m_3-\frac{1}{4} m_2^2 
\eq
The original variables $p_0,r_2,q_1$ can be written as
\ba
p_0 &&= v^{-4/3} \left( m_1 f_1^2 + m_2 f_1 f_2 + m_3 f_2^2 \right) \nonumber \\
r_2 &&=  v^{-4/3} \left(m_1  \dot{f_1} ^2 +m_2   \dot{f}_1 \dot{f}_2 + m_3  \dot{f_2} ^2 \right) \nonumber \\
q_1&&= - v^{-4/3} \left( m_1  f_1 \dot{f}_1 +m_3 f_2 \dot{f}_2 +\frac{1}{2} m_2 (\dot{f}_1 f_2+\dot{f_2} f_1)  \right).
\ea
where now the constraint is given by:
\bq
v^{2/3} \left( p_0 r_2 - q_1^2 \right)  = m_1 m_3-\frac{1}{4} m_2^2 
\eq

In terms of this solution we have Eq. (\ref{aeq})
\bq 
a_{11}=p_0 (t); ~a_{01}=q_0 (t)+q_1 (t) x;~ a_{00} =r_0 (t)+r_1 (t)x+r_2 (t)x^2.
\eq
The metric is given by Eq. (\ref{gij} ).

We find that $ |Det(a)|$ can be written as:
\ba
&&|Det(a)|=\frac{ 1} {4  v ^{2/3}}  \times D \nonumber \\
&&D=\left[n_1 x^2 + x(n_2 f_1+ n_3 f_2)+ 4( n_4 f_1^2 +  n_6 f_2^2+n_5 f_1f_2) \right]. 
 \ea
where
\ba
&&n_1=-m_2^2+4m_1m_3,~ n_2=-4m_2 \alpha+ 8 m_1 \beta, ~n_3 = -8 m_3 \alpha +4 m_2 \beta  \nonumber \\
&&n_4 =  m_1 \br_0 - \alpha^2,  ~n_5=m_2 \br_0  -2 \alpha \beta, ~n_6= m_3 \br_0 - \beta^2.
\ea

From this we find that the affine connection components are in general of the form
\bq 
\frac{f(i,j,k)}{v(t) D}
\eq
where $f(i,j,k)$ is linear in the $n_i$, quadratic in $x$  and also quadratic in the variables $ \left( {f}_i , \dot{f}_i,  \ddot{f}_i \right)$.
The gauge invariant components obey Eq. (\ref{gauge}).

The Ricci Scalar is given by:
\ba \label{ricci3}
R&&=- \frac{1}{2}m_2^2 \br_0 + 2 m_1 m_3 \br_0 - 2 m_3 \alpha^2 + 
 2 m_2 \alpha \beta - 2 m_1 \beta^2  \nonumber \\
 &&= \frac{1}{2} \br_0 n_1 + \frac{1}{4} (n_3 \alpha-n_2 \beta).
 \ea
In case B, we can repeat the process of making a coordinate transformation to eliminate the dependence on one of the functions. Unlike in case A, the functions $f_1 (t)$ and $f_2 (t)$ are on equal footing, so it is equivalent to choose either one. Taking, as before, $f_1 (t)=T-T_0$,  we can proceed to re-express the metric coefficients as functions of $f_2 (T), f'_2 (T), x$ and  $T$. Here too, the Ricci scalar is a constant, this time given in equation Eq. (\ref{ricci3}).
\section{Conclusions}
The investigations reported in this paper were motivated by an attempt to interpret the non-linear extension of quantum mechanics introduced in a previous work. We have assumed that one of the two state vectors represents the system in a particular region of space, such as region I in the spacetime discussed in section V. Then the other state vector should represent the system in region III. 
To see how this might work, we compare the trajectories generated by the two state vectors. In Eq. (\ref{traj}), we represented the trajectory associated with $\kpsi $ as 
\bq \label{traj2} 
X_{\psi} (t)= \frac{2}{\Delta}  \left[(k_1 t+k_2 )\cosh (b y)+ (k_3 t+k_4 ) \sinh (b y) +(k_5 t+k_6 )  \cos (a y)+(k_7 t+k_8 ) \sin (a y) \right].
\eq
Here  $y=2 \omega_0 t$ and $\Delta =N\cosh (b y)+ 2 \omega_0   \sinh (b y) $. 

Using Eq. (1.5), we can perform a similar calculation for the trajectory associated with $\kphi$ to obtain
\bq \label{traj2} 
X_{\phi} (t)= \frac{2}{\Delta'}  \left[(k_1 t+k_2 )\cosh (b y)- (k_3 t+k_4 ) \sinh (b y) -(k_5 t+k_6 )  \cos (a y)-(k_7 t+k_8 ) \sin (a y) \right].
\eq

where now
\bq
 \Delta' = N\cosh (b y) -2 \omega_0   \sinh (b y).
 \eq 
So we have
\bq
X_{\psi} (-k_1,k_2,-k_3,,k_4,k_5,-k_6,-k_7,k_8; -t)=X_{\phi} (k_1,k_2,k_3,,k_4,k_5,k_6,k_7,k_8; t).
\eq
In the simple 2D examples that we have considered, we have sought geodesics in which the parameters analogous to the $k_i$ are freely variable, in which case the metrics derived from  $X_{\chi}$ would be the same as those derived from $X_{\psi}$, once the substitution $t \rightarrow -t $ is made. There would then be no obstacle to interpreting $\kpsi$  and $\kphi$ as representing the system in two disjoint regions, with quantum-mechanical time (i.e. the time appearing in eqs. (1.1) and (1.2)) flowing in opposite directions in the two regions.  This is a quite different scenario from what was discussed in the work of Aharonov and collaborators, \cite{Aharonov1} \cite{Aharonov2}  in trying
to use a two component \Schrodinger\ equation to have a time symmetric Quantum mechanics (see also \cite{gell-mann}).

Of course, the limited investigations we have done, in two dimensional spacetime, do not address the issue of whether our interpretation will survive in more complicated situations. To gain further insight, it will be necessary to pursue the same set of ideas in higher dimensions, where Einstein's equations have dynamical content.

The reader may wonder how our proposed extension of quantum mechanics relates to the scheme introduced by Weinberg \cite{Weinberg}. . In both cases the intent is to explore the possibility of adding non-linear terms to the usual linear quantum mechanics, but our approach is not a special case of Weinberg's scheme. Unlike Weinberg, we introduce a pair of state vectors, and furthermore Weinberg imposes the constraint that the norm of his wave function is constant in time. In our case, the individual norms $<\psi|\psi> $ and $<\phi|\phi >$ vary in time, although the sum of the two does remain constant.


The take-away message of our work is that the non-linear extension of quantum mechanics induces a modification in the time-dependence of the expectation values of operators, in particular position operators, which, in the classical regime, represent the world-lines of the associated particles. If we further assume that, for free particles, these trajectories are geodesics of some metric, then we are led to imagine that gravity is a manifestation of the underlying non-linearity of quantum mechanics. \footnote{  In a totally different context using the functional \Schrodinger\ equation in a gauge theory of nonlinear quantum mechanics, H-T Elze  made the assertion  ``gravity, in this picture, appears as a manifestation of the nonlinearity of quantum mechanics" \cite{Elze}. } 
 These geodesics are the actual observables of gravity. In most, if not all, situations, the metric itself is a quantity inferred from the behavior of particles that are assumed to travel on geodesics.

To pursue this idea further, we need to extend our investigations beyond two dimensions. It will  also be useful to probe more deeply into the meaning of the non-linear extension introduced in \cite{us23}  and perhaps find interesting generalizations thereof. In particular, in \cite{us23} we did not succeed in exhibiting a variational principle from which our equations, (\ref{eq1.1}) and (\ref{eq1.2}), could be derived when $g \neq g*$ (we were able to derive them by introducing a dissipation function, but it is not clear if that was necessary). A variational principle that did not rely on a dissipation function would provide additional understanding, and new ways to analyze the consequences of what we have done.

\acknowledgements
We would like to acknowledge useful conversations with Paul Anderson and Vladimir Matveev. 
We would also like to thank Prof. David Klein  for the use of his diagram of Kruskal coordinates.


\begin{thebibliography}{100} 
\bibitem{us23} A. Chodos, F. Cooper. ``A Solvable Model of a nonlinear extension of Quantum Mechanics", Physica Scripta 98 (4), 045227 (2023).
\bibitem{Jackiw} R. Jackiw, ``Lower Dimensional Gravity," Nucl. Phys. B252 (1985) 343.
\bibitem{Teitelboim} C. Teitelboim, ``Gravitation and Hamiltonian Structure in Two Space-Time Dimensions", Phys. Lett. 126B (1983) 41
\bibitem{Kaplan}  D. E. Kaplan and S. Rajendran, ``Causal framework for nonlinear quantum mechanics".  Phys. Rev. D 105, 055002 (2022)
 \bibitem{Matveev} Vladimir S. Matveev ``Geodesically equivalent metrics in general relativity" , J. Geom. Phys. 62,  675  (2012).
\bibitem{Liouville}  R. Liouville, ``Sur les invariants de certaines equations differentielles et sur leurs applications", Journal de l' Ecole Polytechnique 59 (1889), 7.
  \bibitem{Aharonov1} Yakir Aharonov, Peter G. Bergmann, and Joel L. Lebowitz, ``Time Symmetry in the Quantum Process of Measurement".  Phys. Rev. 134, B1410 (1964).
\bibitem{Aharonov2} Yakir Aharonov, Lev Vaidman``The Two-State Vector Formalism: An Updated Review"   arXiv:quant-ph/0105101v2 (2007). 
\bibitem{gell-mann} M. Gell-Mann and J.B.Hartle, ``Time Symmetry and Asymmetry in Quantum Mechanics and Quantum Cosmology", in ``Physical Origins of Time Asymmetry", ed by J. Halliwell, J. Perez-Mercader, and W. Zurek, Cambridge University Press, Cambridge,(1994).


\bibitem{Bars}  I. J. Araya, I. Bars and A. James, ``Journey beyond the Schwarzschild black hole singularity'' arXiv: 1510.03396  (2015). 
\bibitem{Fuller and Wheeler} Robert W. Fuller and John A. Wheeler
Phys. Rev. 128, 919  (1962).

\bibitem{Weinberg} S. Weinberg, ``Testing Quantum Mechanics" , Annals of Physics 194, 336-386 (1989)).
\bibitem{Elze} Hans-Thomas Elze  ``A relativistic gauge theory of nonlinear quantum mechanics and Newtonian gravity", 	Int.J.Theor.Phys.47:455, (2008)  arXiv:0704.2683 [gr-qc]. 
\end{thebibliography}
\end{document}